\documentstyle[11pt,newpasp,twoside,epsf]{article}
\def\au{\rm AU}
\def\mjup{M_{\rm Jup}}
\def\kms{{\rm km~s^{-1}}}
\def\msun{M_\odot}

\def\edcomment#1{\iffalse\marginpar{\raggedright\sl#1\/}\else\relax\fi}
\reversemarginpar

\begin{document}

\title{Microlensing Constraints on Low-Mass Companions}
 \author{B. Scott Gaudi}
\affil{Hubble Fellow, School of Natural Sciences, Institute for Advanced Study, Einstein Drive, Princeton, NJ, 08540}

\begin{abstract}
Microlensing is sensitive to binary, brown dwarf (BD), and planetary
companions to normal stars in the Galactic bulge with separations
between about 1-10 AU.  The accurate, densely-sampled photometry of
microlensing events needed to detect planetary companions has
been achieved by several follow-up collaborations.  Detailed analysis
of microlensing events toward the bulge demonstrates that less than
45\% of M-dwarfs in the bulge have $\mjup$ companions between 1 and 5
AU.  Detection of binary and BD companions using microlensing is
considerably easier; however, the interpretation is hampered by their
non-perturbative influence on the parent lightcurve.  I demonstrate that $\sim 25\%$ of BD
companions with separations $1-10{\rm AU}$ should be detectable with
survey-quality data ($\sim 1~{\rm day}$ sampling and $\sim 5\%$
photometry).  Survey data is more amenable to generic, brute-force
analysis methods and less prone to selection biases.  An analysis of
the $\sim 1500$ microlensing events detected by OGLE-III in the next
three years should test whether the BD desert exists at separations
$1-10{\rm AU}$ from M-dwarfs in the Galactic bulge.
\end{abstract}

\section{Introduction}

It is by now well-established by radial velocity (RV) surveys that the
mass function of low-mass companions to normal (GKM) stars in the
solar neighborhood exhibits a ``brown-dwarf (BD) desert,'' a paucity
of $13-80\mjup$ companions.  Specifically, $<1\%$ of
solar-type stars have BD companions with semi-major axes
$a\la 5\au$ (Marcy \& Butler 2000).  Although RV surveys will
eventually be able to detect more distant companions, the
duration of their observations are currently only sufficient to detect
long-term trends from companions with $a\ga 5\au$ (See Figure 1).
Young BD companions with separations greater than a few tens
of AUs and less than a few hundreds of AUs can be detected via direct
imaging surveys (Oppenheimer et~al.\ 2001, Lowrance 2001, McCarthy
2001); even more distant companions are detectable serendipitously in
wide-field surveys (Gizis et~al.\ 2001).  These studies seem to
suggest that the frequency of wide BD companions to GKM stars
is $\la 10\%$ (see McCarthy 2001).  Neither RV
nor direct imaging can currently assess the frequency of companions with
$a=5-10\au$.  The exquisite precision ($\sim 3~\kms$) of RV surveys also
makes them sensitive to planetary companions with
mass $0.1-13\mjup$; currently nearly 100 such companions are now
known.  The $M\sin i$ and $a$ of these companions 
are shown in Figure
1.  There also appears to be a statistically significant dearth of
high-mass ($M>5\mjup$), close-in ($a\la 0.3\au$) planetary companions
(Zucker \& Mazeh 2002; see also Figure 1).  Whether this is related to
the BD desert is not clear.

Microlensing is sensitive to binary, BD, and planetary companions separated
by $\sim 1-10\au$ from 
the objects in the Galactic bulge which serve as the lenses of
microlensing events detected toward the Galactic bulge (Mao \&
Paczy\'nski 1991).  The majority of the primaries are likely to be
M-dwarfs in the bulge, with some contamination from stellar remnants
and disk stars.  Microlensing is sensitive to the mass ratio $q$ and
projected separation $d$ of binary systems, where $d$ is in units of
the combined Einstein ring radius $r_{\rm E}$ of the system.  I will
assume that the primaries are M-dwarfs with $M=0.3\msun$ in the bulge,
so that $r_{\rm E}=2\au$, and use these to transform to the mass $M_p$
and semi-major axis $a$ of the secondaries.

\section{Microlensing Constraints on Planetary Companions}

\begin{figure}
\plotone{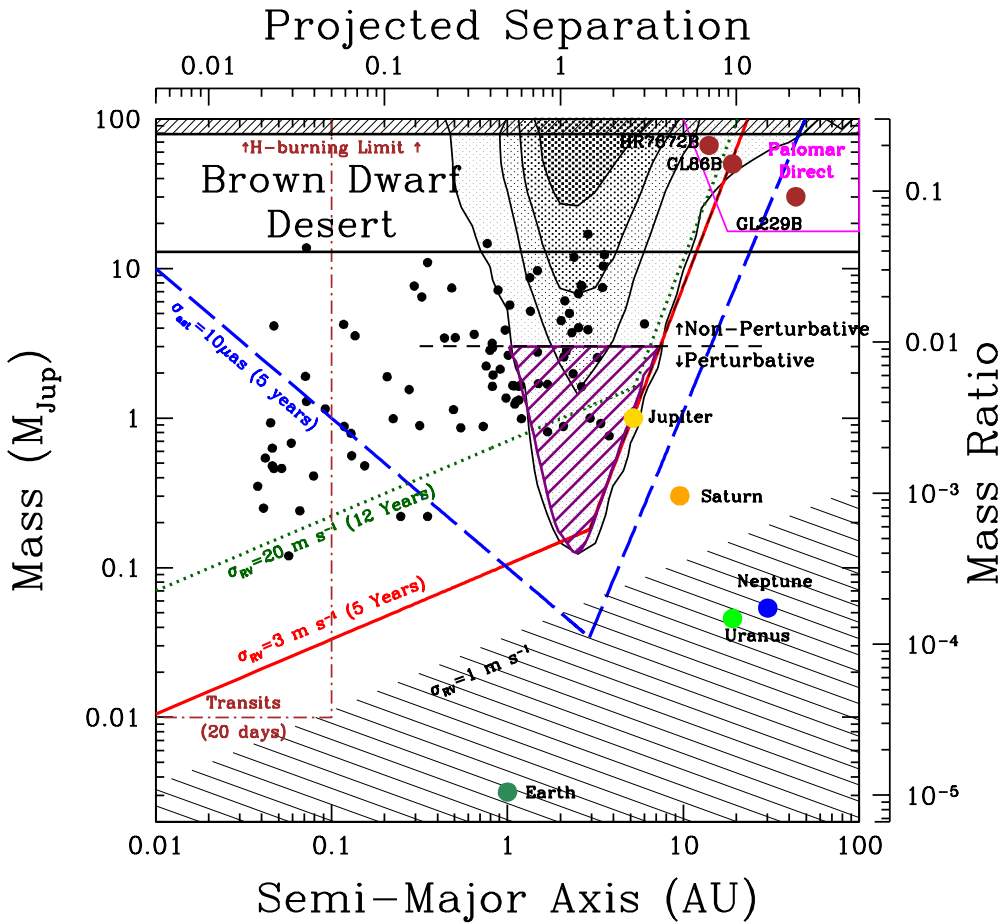}
\caption{Census of brown dwarf and planetary companions to solar-type
stars, as well as sensitivities of various detection methods.  
The small points are companions detected in radial velocity (RV) surveys; the
detection limits of such surveys are shown for the durations and
precisions indicated, assuming solar-mass primaries.  
The hatched
region in the lower right is inaccessible to RV
surveys. The detection limit for an astrometric
survey of solar-mass primaries at 10pc is shown as the long-dashed
line.  The dot-dashed line shows the approximate limit for a 20 day
transit survey.  
The exclusion region from the PLANET microlensing
survey is shown as the triangular hatched region; $<45\%$ of M-dwarfs
in the bulge have companions in this region.  
The region of
sensitivity of the Palomar search is shown as the
quadrilateral in the top right corner.  
The contours show microlensing
detection probabilities of $2.5\%, 25\%, 50\%, 75\%$ (outer to inner)
assuming daily sampling and 5\% photometry.
}
\end{figure}

The collaborations that survey the Galactic bulge generally have
insufficient sampling ($\sim 1~{\rm day}$) to detect the deviations
caused by planetary companions to the primary lenses, which last for a
fraction $\sim \sqrt{q}$ of the total event duration, or $\sim 1~{\rm
day}(M_p/\mjup)^{1/2}$ for typical parameters.  However, these surveys
reduce their data real-time, allowing them to issue alerts of ongoing events.  
Follow-up collaborations can then
monitor only promising events and optimize exposure
times and sampling on an event-by-event basis.  Several such
collaborations have demonstrated that the precision and sampling
needed to detect Jupiter-mass companions is readily achievable (Rhie
et~al.\ 2000; Albrow et~al.\ 2001; Tsapras et~al.\ 2001, Bond et~al.\
2002).

Analyzing events for the presence of planetary signals is made
considerably easier by the perturbative nature of the planetary signal
for $q\la 0.01$: such events are well-described by short-duration
deviations atop otherwise normal (single-lens) lightcurves.
Therefore, selecting the sample of events suitable for analysis is
straightforward, and searching for planetary signals and determining
detection efficiencies is also considerably less complicated and
time-consuming.  A detailed analysis of 43 events monitored by the PLANET collaboration
uncovered no candidate planetary signals.  This null result was used to place an upper
limit to the fraction of lenses that hosted companions.  Figure 1 shows the region of
parameter space in which $<45\%$ of M-dwarfs in the bulge can have
companions at 95\% confidence (Albrow et~al.\ 2001, Gaudi et~al.\
2002).

\section{Microlensing Constraints on Brown Dwarf Companions}

For $q\ga 0.01$, the perturbative nature of the companion begins to
break down, and the resulting lightcurves begin to exhibit
considerably more complexity.  This
introduces several complications to the analysis.  First, the relation
between the lightcurve parameters and the observable features 
is no longer straightforward.  This means that it is
difficult to estimate parameters without detailed fitting;
however detailed fitting is generally quite difficult because of the
highly pathological nature of the $\chi^2$ surface.  These
difficulties are exacerbated in densely-sampled, high-quality
lightcurves, which tend to sample high-magnification portions of the
lightcurve, whose precise shape depend sensitively on minute variations in 
the underlying parameters.  Furthermore, since a large portion of such
lightcurves will appear anomalous, they will generally be recognized
early on, and thus preferentially followed by the follow-up
collaborations.  This introduces hard-to-quantify selection biases.
Therefore, it is likely to be considerably easier to study $q\ga 0.01$
companions with survey photometry, since 
it will be more amenable to global brute-force analysis
methods and less-prone to selection biases (Jaroszy\'nski 2002).

The number of events that must be analyzed to give interesting
constraints on BD companions depends on the sensitivity of
survey-quality photometry.  I estimate this by determining the detection
efficiency as a function of ($M_p,a$), 
assuming uniform daily sampling from 60 days before
until 60 days after the peak, an event timescale of 20 days, a
photometric accuracy of $5\%$, and a uniform distribution of impact
parameters up to $0.8$.  The resulting contours of constant detection
efficiency are shown in Figure 1.  I find that $\sim 25\%$ of BD
$(M_p=13-80\mjup$) companions with separations $1-10\au$ are
detectable with survey photometry.  Therefore, in $N$ microlensing
events, the number of BD detections is approximately $N_{\rm BD}=0.2
f_{\rm BD} N$, where $f_{\rm BD}$ is the frequency of BDs with
$a=1-10\au$.  OGLE-III (http://www.astrouw.edu.pl/$\sim$ftp/ogle/) should detect $\sim 500$ events per year (Udalski et~al.\ 2002);
therefore fractions of $f_{\rm BD} \la 1\%$ can be probed in the next
few years.

\acknowledgements

Support for this work was provided by NASA through a Hubble Fellowship grant
from the Space Telescope Science Institute, which is operated by the
Association of Universities for Research in Astronomy, Inc., under
NASA contract NAS5-26555.

\end{document}